# STATISTICAL MODEL FOR METEOROLOGICAL ELEMENTS BASED ON LOCAL RADIOSONDE MEASUREMENTS IN MEDITERRANEAN REGION

A. Virtser,[1] I. Kupershmidt,[1] Yu.M. Shtemler,[2]
[1]Institute for Industrial Mathematics, Beer-Sheva, 84500, Israel
[2]Department of Mechanical Engineering Ben-Gurion University of the Negev, P.O.B. 653, Beer-Sheva 84105, Israel, shtemler@bgu.ac.il

**Abstract.** A comprehensive statistical model is developed for vertical profiles of the horizontal wind and temperature throughout the troposphere based on several-years radiosonde measurements of strong winds. The profiles measured under quite different atmospheric conditions exhibit qualitative similarity. A proper choice of the reference scales for the wind, temperature and altitude levels allow us to consider the measurement data as realizations of a random process with universal characteristics: means, the basic functions and parameters of standard distributions for transform coefficients of the Principal Component Analysis. The features of the atmospheric conditions are described by statistical characteristics of the wind-temperature ensemble of dimensional reference scales. The model can be useful for air pollution and safety in high-risk areas such as chemical and nuclear plants.

**1. Introduction.** Global Positioning Atmospheric Sounding Systems combining a radiosonde release with tracking the radiosonde provide for the vertical profiles of horizontal wind and temperature as functions of altitude throughout the troposphere. Winds are commonly treated as horizontal flows because maximal horizontal winds are typically in tens of meters per second, while vertical winds are of the order of one meter per second and can reach order of the horizontal winds only in emergencies such as thunderstorms. Dryden's and Karman's standard models for turbulent winds are widely used in engineering.[1] Advanced models subdivide the data into uniform portions of the wind coming from a certain sector of directions during a certain period of the year. The standard Weibull model is applied individually to each portion. Parameters of the advanced models are fitted to histograms of measurement data.[2] The input parameters of the models are suitable for local applications within the atmospheric boundary layer, while they are poorly defined at wide range of parameters. Goal of the present study is development of statistical modeling of vertical-wind profiles throughout the troposphere. Key idea is the data homogenization by a proper scaling.

**2. Preliminary processing of 3-year radiosonde measurements of wind and temperature.** Altitude levels of the maximum wind, its direction and temperature are highly expressed for strong winds important in practice (~70% of all winds observed are strong). Maximal-wind rose for strong winds over the 3-year period is shown in Fig. 2. Strong winds have the west-sound prevailing direction (about 85% of their total number). It is seen that strong winds have the west-sound prevailing direction (about 85% of their total number). Seasonal variations of reference scales over 3-year meteorological data are depicted in Fig. 3. Reference scales vs date of measurement. Day and evening measurements are made daily (two ordinates correspond to every abscissa)

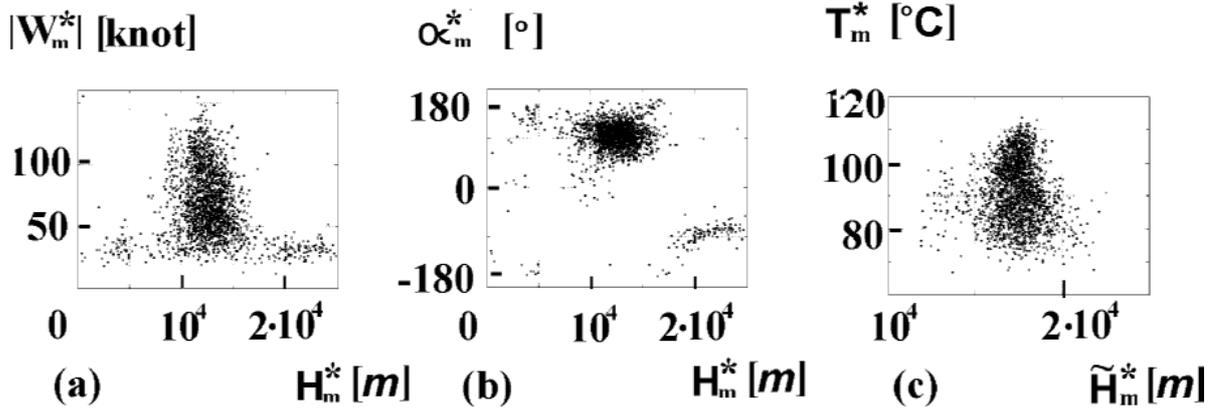

Figure 1. Reference scale distributions over the 3-year period (local measurements, Mediterranean region), $W_m^* = W_m^{max}$, $\alpha_m^* = \alpha_m^{max}$, $H_m^* = H_m^{max}$; $T_m^* = |T_m^{inv} - T_m^{(0)}|$, $\widetilde{H}_m^* = H_m^{inv}$.

Reference wind speed (a) and direction (b) vs altitude level of maximal wind-speed.

Reference temperature (c) vs altitude level of inverse temperature.

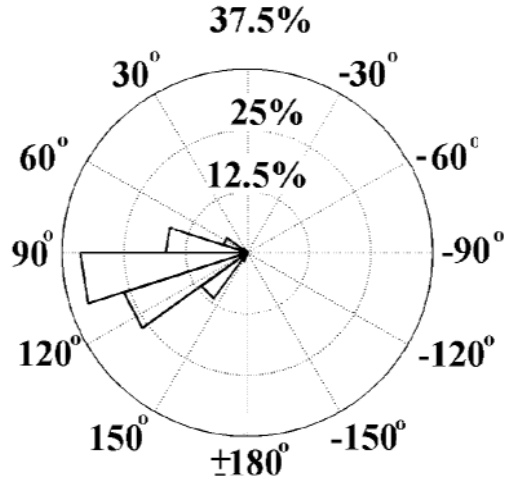

Figure 2. Strong wind rose for over the 3-year period

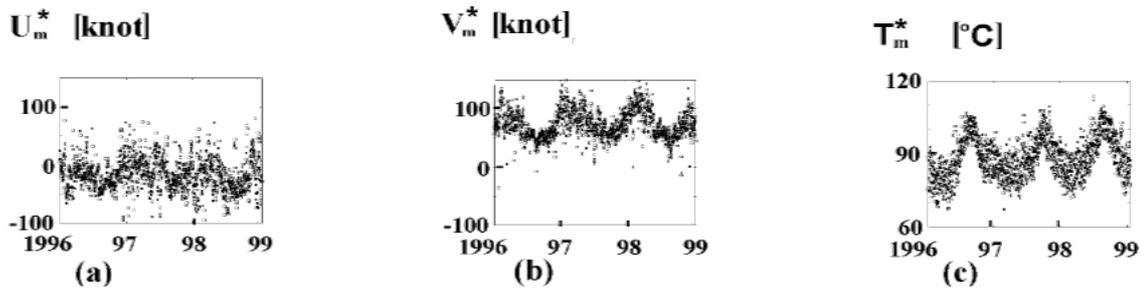

Figure 3. Seasonal variations of reference scales of 2 component wind speed and temperature over 3-year meteorological data.

**3. Homogenization of Meteorological Data.** Wind-vector scaling and rotating as well as temperature scaling are applied to homogenize meteorological data (see vertical profiles of wind and temperature before and after homogenization in Fig. 4):

$$\mathbf{w}_m(h) \equiv \frac{\hat{\mathbf{W}}_m(H)}{W_m^{max}} = \Omega \frac{\mathbf{W}_m(H)}{W_m^{max}}, \quad t_m(\widetilde{h}) \equiv \frac{\Delta T_m}{\Delta T_m^{max}} = \frac{T_m - T_m^{(0)}}{T_m^{inv} - T_m^{(0)}}$$

where

$$\Omega = \begin{pmatrix} \cos\alpha_m^{max} & \sin\alpha_m^{max} \\ -\sin\alpha_m^{max} & \cos\alpha_m^{max} \end{pmatrix}, \quad h = \frac{H}{H_m^{max}}, \quad \tilde{h} = \frac{H}{\tilde{H}_m^{max}}$$

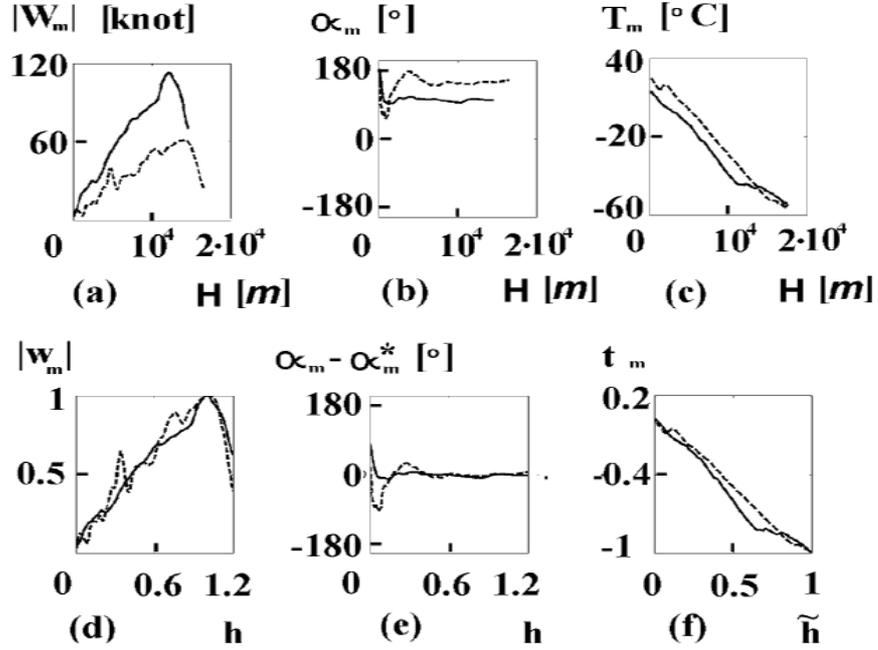

Figure 4. Vertical profiles of wind and temperature before(a-c) and after homogenization(d-f)

**4. Principal Component Analysis.**[3] Mean wind and temperature, as well as PCA expansions for deviations of wind and temperature are as follows:

$$\overline{\mathbf{w}}(h) = \frac{1}{M}\sum_{m=1}^{M}\mathbf{w}_m(h), \quad \bar{t}(\tilde{h}) = \frac{1}{M}\sum_{m=1}^{M}t_m(\tilde{h})$$

$$\mathbf{w}'_m(h) \equiv \mathbf{w}_m(h) - \overline{\mathbf{w}}(h) = \sum_{k=1}^{2K} c_m^{(k)}\mathbf{v}^{(k)}(h) + ... + \varepsilon_m(h),$$

$$t'_m(\tilde{h}) \equiv t_m(\tilde{h}) - \bar{t}(\tilde{h}) = \sum_{k=1}^{K} \tilde{c}_m^{(k)}\tau^{(k)}(\tilde{h}) + ... + \tilde{\varepsilon}_m(\tilde{h})$$

Square errors vs total number of basic functions are shown in Fig. 5. Dimensionless means, basic functions and standard deviations are presented in Fig. 6.

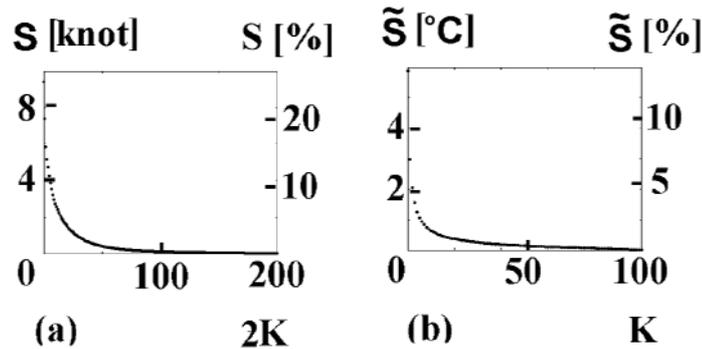

Figure 5. Square errors vs total number of basic functions

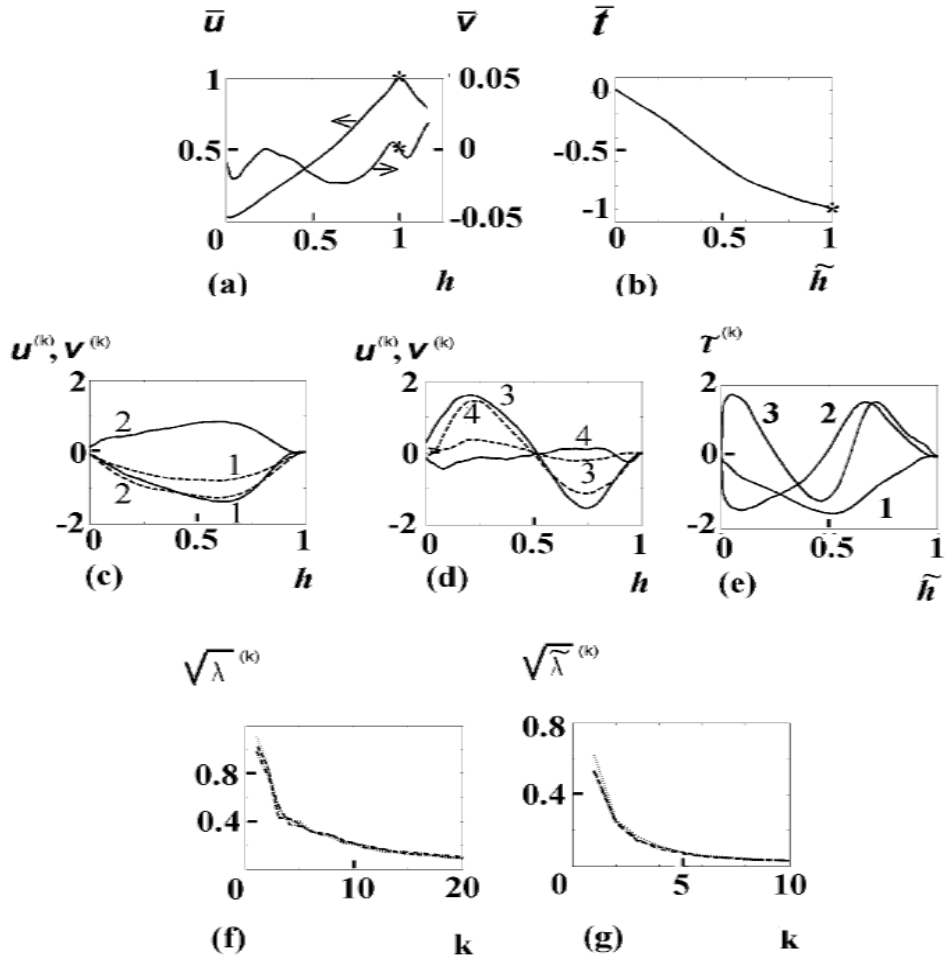

Figure 6. Dimensionless means, basic functions and standard deviations.

(a) Mean wind speed vs $h$, (b) Mean temperature vs. $\tilde{h}$

(c)- (e) wind and temperature basic eigenfunctions

(f)- (g) - standard deviations for PCA model coefficients of wind and temperature vs. current number of modes $k$.

Note that the value K=0 in in the above relations correspond to approximations of the mean wind and temperature. In accordance with Fig. 6, dimensional mean-square errors for the mean wind and temperature are 9.5 knots and 5.9 °C (K =0), 3-mode approximation (K =3) yields 5 knots and 2 °C, and 5-mode approximation (K =5) monotonically reduces the dimensional mean-square errors to 2.5 knots and 1°C. In accordance with available experimental data the micro-scale components of wind and vertical winds that are beyond our consideration can be of the order of 2-5 knots. So, although any required exactness can be achieved, too high exactness has little meaning, and physically admissible values of the wind mean-square error correspond to 5- or even 3-mode approximation. Note that at K=3, k-mode approximation corresponds to the wave length of the order of $\bar{H}_m^*/K \sim 4 \cdot 10^3$ m in the vertical direction

**5. Statistical Modeling of Local Meteorology.** The following main hypothesa are adopted. Two ensembles (dimensional & dimensionless parameters) are independent and normal: (i) PCA model coefficients with standard deviations (ii) Reference scales with the correlation matrix (see the Table 1). Thus, the wind speed and temperature are transformed as follows:

$$\mathbf{W}(H) = W^{max}\Omega(\alpha^{max}) \left[\overline{\mathbf{w}}\left(\frac{H}{H^{max}}\right) + \sum_{k=1}^{2K} c^{(k)}\mathbf{v}^{(k)}\left(\frac{H}{H^{max}}\right)\right],$$

$$T(H) = T(0) + \Delta T^{max}\left[\bar{t}\left(\frac{H}{\widetilde{H}^{max}}\right) + \sum_{k=1}^{K} \widetilde{c}^{(k)}\tau^{(k)}\left(\frac{H}{\widetilde{H}^{max}}\right)\right].$$

|  | $H^{max}$ | $U^{max}$ | $V^{max}$ | $\widetilde{H}^{max}$ | $\Delta T^{max}$ | $T(0)$ |
|---|---|---|---|---|---|---|
| $H^{max}$ | 1 | 0.04 | -0.09 | 0.03 | 0.06 | 0.05 |
| $U^{max}$ | 0.04 | 1 | 0.09 | -0.03 | -0.17 | -0.17 |
| $V^{max}$ | -0.09 | 0.09 | 1 | 0.02 | -0.45 | -0.47 |
| $\widetilde{H}^{max}$ | 0.03 | -0.03 | 0.02 | 1 | -0.14 | -0.15 |
| $\Delta T^{max}$ | 0.06 | -0.17 | -0.45 | -0.14 | 1 | 0.93 |
| $T(0)$ | 0.05 | -0.17 | -0.47 | -0.15 | 0.93 | 1 |

Table 1. Matrix of Correlation Coefficients for Reference Scales

Measured and simulated vertical profiles and their PCA approximations are compared in Fig.7.

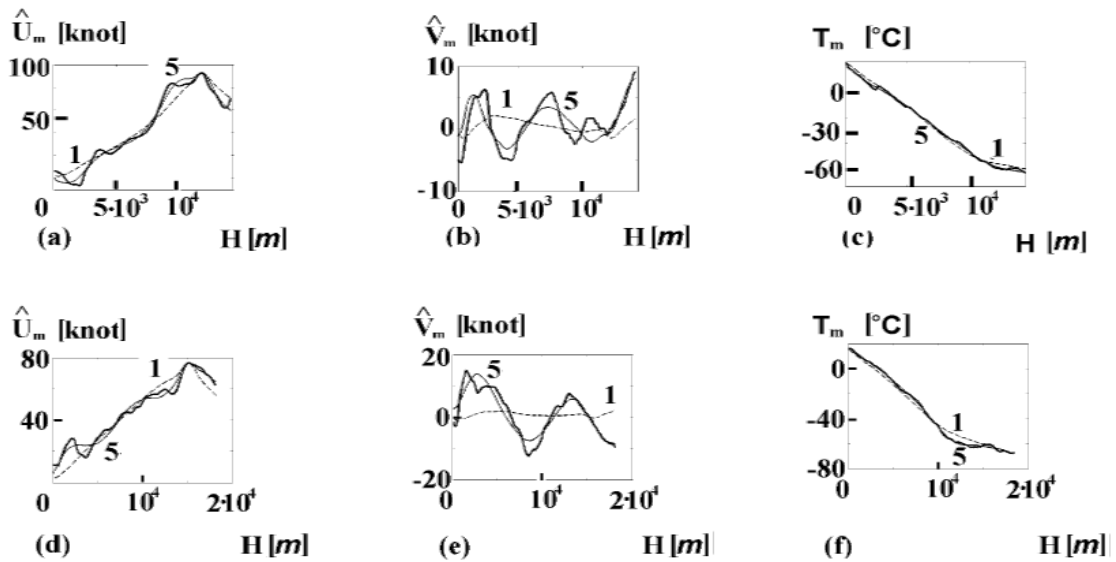

Figure 7. Measured (a-c), simulated (d-f) vertical profiles (bold curves) and their
PCA approximations at $K=1$ and 5 functions (thin and dash curves).
Longitudinal (a), (d) and transversal (b), (e) components of wind and temperature (c), (f).

**6. Conclusions and discussion.** The joint statistical model of the vertical profiles for local horizontal wind and temperature is developed on the basis of the standard on-site radiosonde meteorological data set in Mediterranean region. For strong winds, i.e. larger 50 *knot,* the altitude level of the maximum wind is highly expressed at $1.0 \cdot 10^4 \div 1.5 \cdot 10^4 m$, and temperature decreases with altitude up to the level $1.3 \cdot 10^4 \div 1.5 \cdot 10^4 m$ where the inverse temperature occurs. The scaling procedure of the rotation by the reference angle along with conversion-scaling factor is developed to homogenize the seasonal & daily variations of meteo-data. PCA is applied to the homogenized wind/temperature data, which are considered as a random process with universal characteristics (means, PCA basic functions and coefficients of PCA transform are independent of atmospheric conditions). Results of statistical modeling are in fair agreement with meteorological data.

The mean wind speed and temperature are found to be near linear with altitude at least for altitudes lower than the altitude level of the maximum wind. Deviations from the mean wind and temperature are described by expansion over eigenfunctions of PCA transform. The model developed yields a fair approximation for observation data with a relatively small mean-square error that vanishes fast with growth of the mode number. Although any required exactness can be achieved, too high exactness has little meaning, and physically admissible values of the wind mean-square error correspond to 5- or even 3-mode approximation. Since the low-mode expansions are valid for the physically meaning approximation of the wind and temperature, the whole-world meteorological data base with relatively rough vertical grid may be used for the present analysis in other regions, where the high-level jets are prevailing. The dimensionless statistical characteristics of the local meteorology calculated independently for three consequent years are found to be very close to one another which demonstrates the annual stability of the meteorological database at least for the main statistical characteristics of the low order modes. The annual characteristics of the higher order modes, however, were subject to significant variations from year to year.

In some processes occurring over extremely short time intervals, the knowledge of the local meteorology at the initial instant is needed. Although the measurements within a given moving radiosonde balloon flight are non-simultaneous (about 1.5-2 hours), meteorological data measured by the radiosonde during its flight can be recalculated to an the instant and place of the balloon start applying, for instance, Taylor's frozen turbulence hypothesis.[4] The frozen turbulence hypothesis allows also to estimate wave characteristics in horizontal plane through the wave characteristics in the vertical direction by implementing the property proportionality of scales in vertical and horizontal directions.[5]